\documentclass{article}
\usepackage{spconf,amsmath,graphicx}
\usepackage{multirow, booktabs}
\usepackage[ 
            colorlinks=True,
            linkcolor=blue
            ]{hyperref}

\title{audio-visual speech separation based on joint feature representation with cross-modal attention}
\name{Junwen Xiong$^{1}$, Peng Zhang$^{1\dagger}$\thanks{$^{\dagger}$Corresponding author: zh0036ng@nwpu.edu.cn}, Lei Xie $^{1}$, Wei Huang$^{2}$, Yufei Zha$^{1}$, Yanning Zhang$^{1}$}

\address{ $^{1}$ASGO, School of Computer Science, Northwestern Polytechnical University, Xi’an, China \\
        $^{2}$China Mobile-NCU AI\&IOT Jointed Lab, Informatization Office, Nanchang University, Nanchang, China}

\begin{document}
%
\maketitle
\begin{abstract}
	Multi-modal based speech separation has exhibited a specific advantage on isolating the 
	target character in multi-talker noisy environments. Unfortunately, most of current separation 
	strategies prefer a straightforward fusion based on feature learning of each single modality, 
	which is far from sufficient consideration of inter-relationships between modalites. Inspired 
	by learning joint feature representations from audio and visual streams with attention mechanism, 
	in this study, a novel cross-modal fusion strategy is proposed to benefit the whole framework with 
	semantic correlations between different modalities. To further improve audio-visual speech separation,
	the dense optical flow of lip motion is incorporated to strengthen the robustness of visual 
	representation. The evaluation of the proposed work is performed on two public audio-visual 
	speech separation benchmark datasets. The overall improvement of the performance has 
	demonstrated that the additional motion network effectively enhances the visual representation 
	of the combined lip images and audio signal, as well as outperforming the baseline in terms of 
	all metrics with the proposed cross-modal fusion.


	
\end{abstract}
%


\section{Introduction}
\label{sec:intro}

\begin{figure}
	\includegraphics[width=0.65\textwidth]{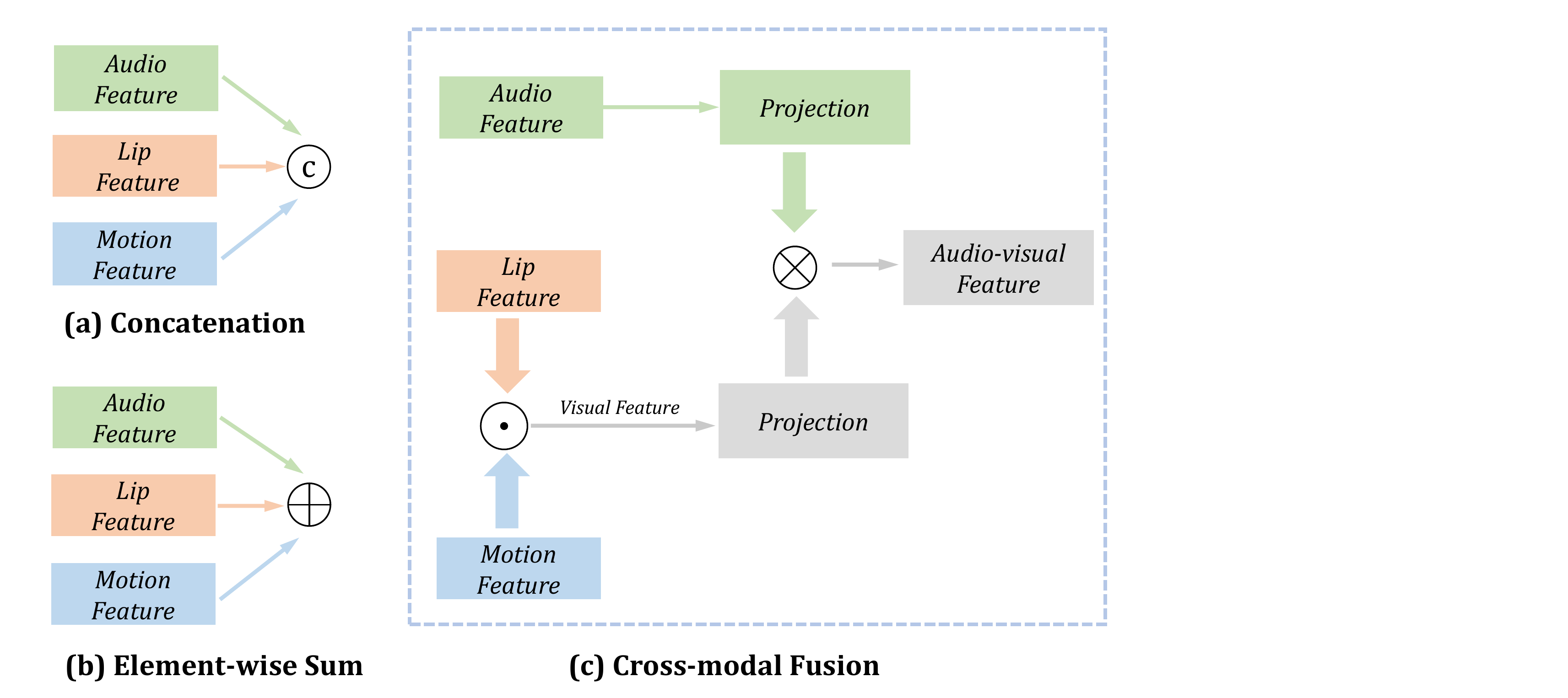}
	
	\vspace{-10pt}
	\caption{The thumbnails of different multi-modal fusion techniques, (a) and (b) show two traditional fusion methods, (c) presents the proposed cross-modal fusion strategy. Operators $\bigodot$ and $\bigotimes$ denote the affine transformation and cross-modal attention respectively.}
	\label{Figure 1}
\end{figure}

\vspace{-10pt}
Nowadays, speech separation is widely studied because of its functionality for different applications such as hearing aid for the elderly and dialogue systems. To effectively discriminate the characteristics of speakers, audio-stream based approaches have been extensively exploited, such as deep clustering \cite{hershey2016deep} and permutation invariant training \cite{yu2017permutation, kolbaek2017multitalker}. Unfortunately, label permutation \cite{hershey2016deep},
a fundamental problem with audio-only speech separation, is difficult to consistently associate the separated audio stream with its corresponding speaker in the mixture signal, especially when the separated signals have similar vocal characteristics \cite{yu2017permutation}.
Another audio-only separation is based on the computational auditory scene analysis (CASA) \cite{hu2012unsupervised}, but its insufficient learning ability limited the overall performance to be further improved yet.

Noticing that in crowded restaurants or loud bars, human perception systems are expertized in understanding the surrounding environments, e.g. to focus primarily on the voices from the interested target, while ignoring to concurrent irrelevant interference at the same time. This capability of speech perceiving under complex conditions not only depends on the human auditory system, but is also benefited from the visual sensing to facilitate the multisensory perception \cite{golumbic2013visual, rahne2007visual}. Subsequently, active speaker detection \cite{tao2021someone}, audio-visual speech separation \cite{gao2021visualvoice}, audio-visual synchronization \cite{morgado2020learning} and etc, have been inspired as the multi-modal based studies.

By analyzing the correspondence between face motions and speech utterances, more efforts have been taken to design multi-modal schemes for effective audio-visual speech separation (AVSS) \cite{gabbay2017visual, ephrat2018looking, afouras2018conversation, lu2018listen}. These methods can automatically separate the audio signal corresponding to the visual part from the mixed sound, just as to determine whether the speaker is silent or not via observing the mouth movement. However, training on small-scale datasets causes the extracted visual features to be easily disturbed when facing more complex scenes.

\begin{figure*}[htp]
	\centering
	\includegraphics[width=\textwidth]{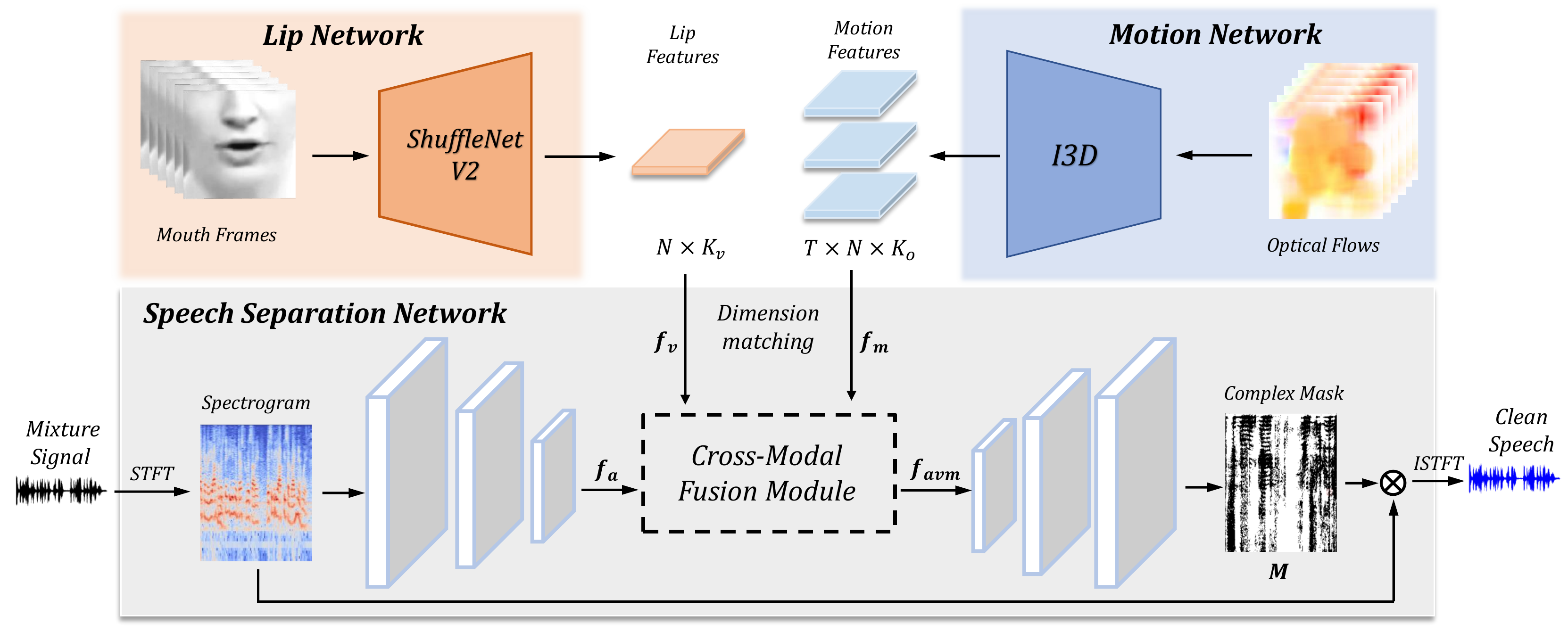}
	\vspace{-25pt}
	\caption{The proposed AVSS framework is composed by 4 parts: a lip network, a motion network, a cross-modal fusion module and a speech separation network. The lip network takes a sequence of mouth regions $x_v$ and outputs lip feature $f_v$; motion network takes the optical flows $x_m$ corresponding to the mouth regions $x_v$ as input and outputs the motion feature $f_m$; cross-modal fusion module integrates
		audio feature $f_a$, which generated from speech separation encoder, lip feature $f_v$ and motion feature $f_m$; speech separation decoder separates clean signal with the fused data $f_{avm}$.}
	\label{Figure 2}
\end{figure*}

To achieve more robust speech separation in diverse scenario, a multi-task modeling strategy \cite{gao2021visualvoice} has been proposed as a promising solution. By learning the cross-modal embedding to establish the matching of face-to-voice \cite{gao2021visualvoice} enables a mutual association to overcome the main challenge of audio-visual inconsistency. Meanwhile, even the straightforward concatenation of different modalities can guarantee the model efficiency, the left room for intrinsic-relationships exploration is still a motivation for successive study.

In this paper, we proposed a novel cross-modal fusion framework to learn joint feature representation from audio and visual information. Besides analyzing the facial attributes, a motion network is designed by incorporating the temporal movement of mouth regions to capture motion cues from optical flow. Considering the complexity of the interaction between different modalities, a cross-modal attention fusion strategy is also proposed to utilize the inter-relationships between audio and visual stream. With these means, a comprehensive solution for audio-visual speech separation by applying the attended value of visual features to the mixed audio features using cross-modal attention.

The effectiveness of the proposed AVSS model is demonstrated on two large-scale public datasets, VoxCeleb2 \cite{chung2018voxceleb2} and LRS2-BBC \cite{afouras2018deep}. The experiment results show that the performance of the proposed work has obtained a significant improvement in comparison to the other state-of-the-art methods. Moreover, the ablation studies indicate that the proposed cross-modal attention is able to benefit the learned joint representations in challenging scenarios.

\vspace{-10pt}
\section{Method}
\label{sec:method}

\vspace{-10pt}
In this section, we first introduce the proposed AVSS framework as shown in the diagram of Fig.\ref{Figure 2}, which presents the leveraging process of lip motion, optical flow and vocal audio for simultaneous separation guidance. Then, the proposed cross-modal fusion strategy is introduced in detail with explaination of the objective function used in the model.


\vspace{-15pt}
\subsection{Network Architectures}
\vspace{-7pt}
The proposed framework adopts the commonly used Mix-and-Separate training paradigm \cite{ephrat2018looking}. The first task is to mix the speech signals of multiple different video clips, and then separate the audio signals from mixture conditioned on visual inputs of the corresponding speaker. In the following, we explicitly introduce the architecture of audio-visual speech separation networks in terms of lip network, motion network and speech separation network.


\noindent{\bfseries {Lip Network}}\quad The lip network performs feature extraction from each input video frame. The the state-of-the-art lip reading model \cite{martinez2020lipreading} is employed as functional structure, which is composed of a 3D convolutional layer and followed by a ShuffleNet v2 network \cite{ma2018shufflenet}. The lip network takes ${N}$ consecutive stacked gray images $x_v$ ($H\times{W}\times{N}$) to generate a lip feature vector $f_v$ of dimension $K_v$.

\noindent{\bfseries{Motion Network}}\quad For the complementary motion cues, it is intended to capture spatial and temporal information with stable performance. Inspired by recent advances in the action recognition studies, a pre-trained I3D model \cite{carreira2017quo} is incorporated into our framework as the motion network. By inflating 2D ConvNets to 3D to obtain an additional temporal dimension, the motion network can extract motion embedding $f_m$ of dimension ${K_o}$ from the optical flows $x_m$ ($H\times{W}\times{(N-1)}\times{2}$) which are estimated by the previous gray images.

Then, the dimension of lip features and motion features is matched before they are fed into the cross-modal fusion module. The matching operation is to multiply the time dimension of the motion feature with the channel dimension, which is to generate the motion features of the same dimension as the lip features.


\noindent{\bfseries{Speech Separation Network}}\quad The architecture of the speech separation network adopts the style of a U-Net \cite{ronneberger2015u}, in order to make the size of output mask to be identical as the input. The network is composed of an encoder and a decoder. The input of the encoder is the complex spectrogram $x_a$ of the mixed signal, which is the 2D time-frequency representation. After the input is processed by a series of convolution layers and pooling layers, the dimension of the complex spectrogram has been compressed, and a well-expressed feature map can be obtained at the same time.

As the input of next decoder, the audio-visual feature $f_{avm}$ is combined by performing cross-modal fusion with the audio feature $f_a$, lip feature $f_v$ and motion feature $f_m$. The decoder output is the predicted complex mask $M$ with the same dimension as the input spectrogram. During inference process, the  speech spectrogram can be predicted by multiplying the estimated complex mask on the input spectrogram, and changed to the final separated speech signal using inverse Short-Time Fourier Transform (iSTFT).


\vspace{-7pt}
\subsection{Cross-Modal Fusion}
\label{sec:fusion}

\vspace{-7pt}
To achieve a joint representation of different modalities, a novel fusion strategy is proposed by introducing 
a cross-modal interaction function. The overall structure of fusion module is shown in Figure \ref{Figure 1}c. 
Before interacting with different data modalities, features of the same modality is firstly fused. Specifically, 
the lip feature and motion feature are processed by applying feature-wise affine transformation which is same as 
the FiLM proposed in\cite{perez2018film}. As shown in the Eq.\ref{eq 1}, FiLM performs affine transformation on 
lip feature ${f_v}$ by the condition of motion feature ${f_m}$ as:


\vspace{-7pt}
\begin{equation}
	\mathbf{FiLM}(f_m, f_v) = \gamma(f_m) \cdot f_v + \beta(f_m)
	\label{eq 1}
\end{equation}

\noindent where ${\gamma(\cdot)}$ and ${\beta(\cdot)}$ are both single fully connected layer which output the scaling vector and bias vector.

The modulated lip feature ${f_{vm}}$ is generated before the process of TCN block \cite{bai2018empirical}, which consists of 1D convolutional layers, batch normalization (BN) and rectified linear unit (ReLU). The function of this module is to capture the temporal relationships in modulated lip features. With all above, the visual feature for joint representation learning is obtained.

\begin{figure}
	\includegraphics[width=0.65\textwidth]{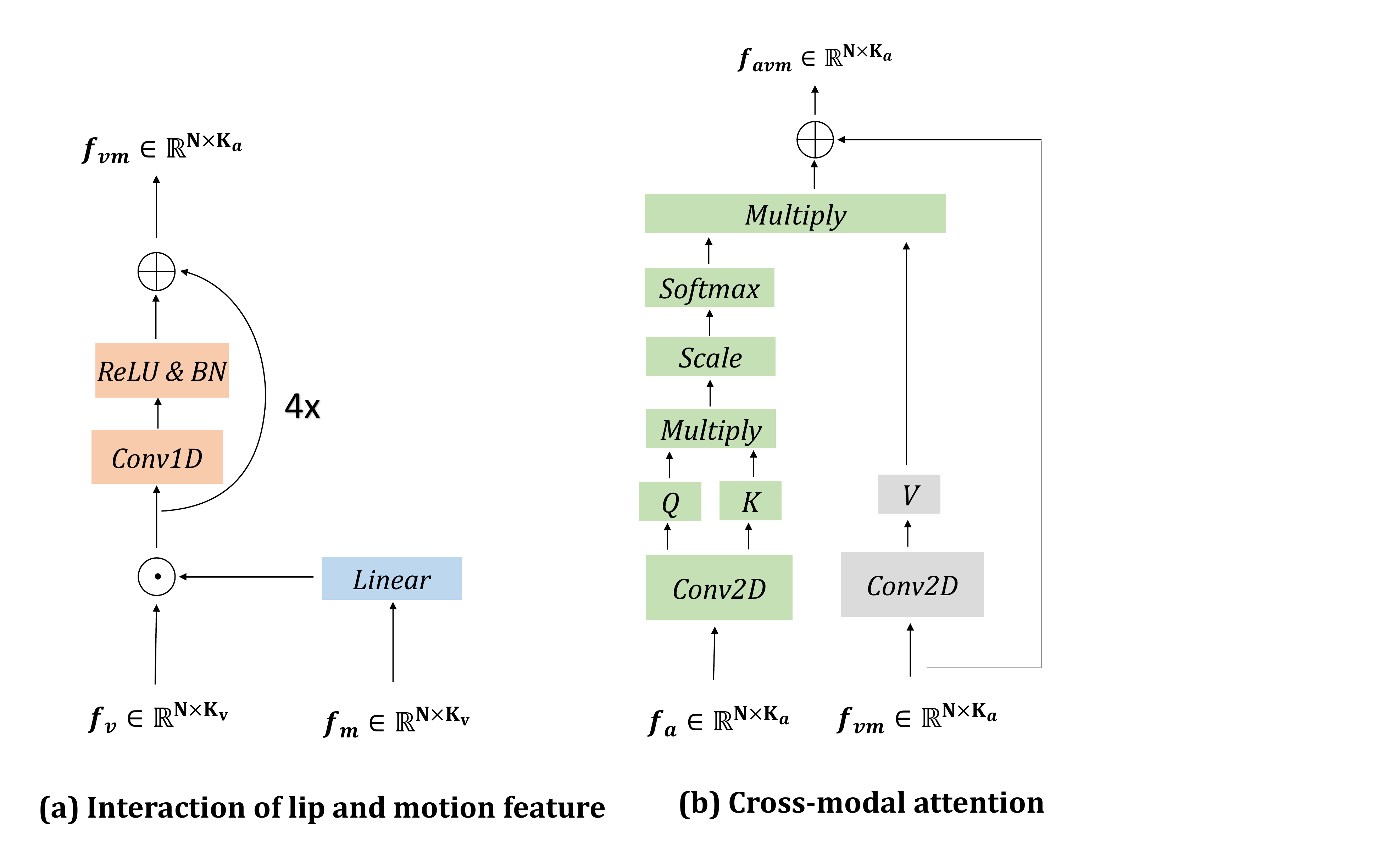}
	\vspace{-25pt}
	\caption{The structure diagram of each part in the fusion module. (a): The interaction includes affine transformation and TCN blocks
		(b): The layer of cross-modal attention.}
	\label{Figure 3}
\end{figure}

The most critical functionality of the fusion network is the cross-modal attention (CMA), as shown in Figure \ref{Figure 3}b. It is inspired from multi-head attention in transformer \cite{vaswani2017attention}, which mainly focuses on discovering the inter-relationships across modalities.

In equation of cross-modal Eq.\ref{eq 2}, we design a learnable parameter $\lambda$ to adaptively adjust the 
attention weight, as well as a residual connection $I(f_m)$ to speed up the model convergence. The input data 
are the query ($Q_{vm}$) and key ($K_{vm}$) from visual feature $f_{vm}$ and the value ($V_{a}$) from audio feature $f_{a}$, 
which have been processed by a 2D convolution layer. The output is the combined feature from audio and visual features.


\vspace{-18pt}

\begin{equation}
	\mathbf{CMA}(f_{vm}, f_a) = \lambda \cdot softmax(\frac{Q_{vm}K_{vm}}{\sqrt{d}})V_{a} + f_{vm}
	\label{eq 2}
\end{equation}

\vspace{-5pt}
where $d$ denotes the dimension of $Q$, $K$ and $V$.

\vspace{-5pt}
\subsection{Training Objective}
\vspace{-5pt}
The predict value of the model is complex mask corresponding to the visual input, shown in the Fig.\ref{Figure 2}. The entire framework is trained by minimizing the $L2$ loss between the predicted complex mask and the ground-truth complex mask.

\begin{equation}
	\mathcal L =\sum_i \left |\right |\hat{M_i}-M_i \left |  \right |_2
\end{equation}

\vspace{-5pt}
where $\hat{M_i}$ denotes the predicted complex mask, and $M_i$ denotes the ground-truth mask, respectively.

\vspace{-10pt}
\section{Experiments}
\label{sec:experiment}

\begin{table*}[!htp]
	\centering
	\begin{tabular}{c|ccccc|ccccc}
		\toprule
		\multicolumn{1}{c|}{ \multirow{2}{*}{Methods} } & \multicolumn{5}{c|}{Seen heard test} & \multicolumn{5}{c}{Unseen unheard test}\\
		\cline{2-11}
		
		&SDR&SIR&SAR&PESQ&STOI & SDR&SIR&SAR&PESQ&STOI\\
		\midrule
		Audio-Only \cite{yu2017permutation} & 7.85& 13.7& 9.97& 2.6& 0.82 & 7.85& 13.7& 9.97& 2.6& 0.82 \\
		AV baseline \cite{gao2021visualvoice} & 8.70&13.98 &10.9 &2.65 &0.82 & 8.64 &13.96&10.71&2.63&0.82 \\
		Ours, Lip & 8.25&13.3&10.5&2.62&0.82 & 8.06&13.2&10.2&2.61&0.81 \\
		Ours, Lip+Flow+Addition & 8.83&14.09&10.86&2.66&0.82 & 8.70&13.90&10.73&2.62&0.82 \\
		Ours, Lip+Flow+Concatenation & 8.85&14.11&11.03&2.67&0.82 & 8.77&14.01&10.80&2.64&0.82 \\
		Ours, Lip+Flow+Cross-Modal & \bfseries{9.19}& \bfseries{14.75}& \bfseries{11.42}& \bfseries{2.76}&\bfseries{0.84} &
		\bfseries{8.90}& \bfseries{14.15}& \bfseries{11.2}& \bfseries{2.67}& \bfseries{0.83} \\
		\bottomrule
		
	\end{tabular}
	\vspace{-5pt}
	\caption{Performance evaluation of speech separation on VoxCeleb2 dataset.}
	\label{table1}
\end{table*}

\vspace{-7pt}
\subsection{Dataset}
\vspace{-5pt}
The proposed model is trained on two large-scale public datasets: LRS2-BBC \cite{afouras2018deep} and VoxCeleb2 \cite{chung2018voxceleb2}. LRS2-BBC dataset contains thousands of sentences and hundreds hours of in-the-wild video clips, which are all from BBC television programs. Another larger dataset VoxCeleb2, contains over 1 million utterances recorded by more than 6000 different speakers, and are collected from Youtube Channels.

The LRS2-BBC dataset has already been divided into different parts of pre-train, train-val and test. The pre-train part is not used in our experiment, and the other two parts are is cataloged according to the broadcast date to avoid overlapping between the sets. There are about 47k utterances in train-val set, and about over 1k in the test set.

The VoxCeleb2 dataset is one of the largest datasets, which is composed of over 2400 hours of audio-visual materials, for 
evaluation of audio-visual speech separation. It is divided into the sets of train-val and test by identity, 
each of which contains 5994 identities and 118 identities. By following the practice of recent work \cite{gao2021visualvoice}, 
two videos are hold out for each identity in the training set as the seen-heard-test set, we divide the original 
test into two parts equally, one as the validation set and the other as the unseen-unheard-test set.

\vspace{-4pt}
\begin{table}[!htp]
	\centering
	\begin{tabular}{cccc}
		\toprule
		Methods& SDR& SAR& PESQ \\
		\hline
		Deep-Clustering \cite{hershey2016deep} & 6.0 & - & 2.3 \\
		AV baseline \cite{gao2021visualvoice}& 10.70& 11.69& 2.73 \\
		Ours (Cross-Modal)& \bfseries{11.04}&  \bfseries{12.07} &\bfseries{2.79} \\
		\bottomrule
	\end{tabular}
	\vspace{-4pt}
	\caption{Evaluation results of speech separation performance on LRS2-BBC dataset.}
	\label{table2}
\end{table}

\vspace{-20pt}
\subsection{Experiment Configurations}
\vspace{-6pt}
The proposed AVSS network is implemented using Pytorch toolkit. By incorporating a motion network to obtain optical flow, the processing pipeline of lip data and audio data is based on \cite{gao2021visualvoice} with preprocessing of training data. Specifically, the optical flow is calculated inside the ROI of mouth region using Gunnar Farneback' algorithm \cite{farneback2003two}.


We use AdamW with $10^{-2}$ weight decay as the network optimizer, the initial learning rate is $10^{-4}$, and decays the learning rate by half
with $8\times 10^4$ of each iteration. All experiments are conducted on 2 RTX-2080Ti GPUs, with batch-size of 8.

\vspace{-15pt}
\subsection{Speech Separation For Different Datasets}
\label{sec:results}

\vspace{-4pt}
\noindent{\bfseries{VoxCeleb2}}\quad Table\ref{table1} reports the results evaluated in both test sets. To simplify the expression, the lip network, motion network and cross-modal fusion abbreviated as Lip, Flow and Cross-Modal, respectively. It can be seen that the proposed model achieves significant performance superior to all the other methods. The ablation experiments also show that the motion net-based model has improved the separation performance by a large margin, which demonstrates that the features of optical flow generated by motion net can enhance visual representation with lip features. In addition, we use concatenation and element-wise sum in the entire framework to replace the cross-modal fusion for experiments. In the compared works, the proposed fusion outperforms all the other fusion methods in both test sets, which indicates that our cross-modal strategy can achieve better joint feature representation.

\vspace{-8pt}
\begin{table}[!htp]
	\centering
	\begin{tabular}{cccc}
		\toprule
		Methods& SDR& SAR& PESQ \\
		\hline
		w/o CMA& 10.94& 11.97& 2.77 \\
		w CMA& \bfseries{11.04}&  \bfseries{12.07} &\bfseries{2.79} \\
		\bottomrule
	\end{tabular}
	\vspace{-4pt}
	\caption{Ablation experiments of cross-modal attention on LRS2-BBC dataset.}
	\label{table3}
\end{table}

\noindent{\bfseries{LRS2-BBC}}\quad Table\ref{table2} reports the comparison between different models on LRS2-BBC dataset. It can be found that the propose model obtains a remarkable improvement with all metrics compared to the previous works \cite{gao2021visualvoice} \cite{hershey2016deep}.Table\ref{table3} shows the ablation experiments for cross-modal attention (CMA) of the proposed fusion. It is obvious that the fusion with cross-modal attention performs better than those without, which also means that the inter-relationship between audio stream and visual stream can be effectively bridge by the proposed fusion with cross-modal attention for speech separation.



\vspace{-10pt}
\section{Conclusions}
\label{sec:conclusion}
\vspace{-5pt}
In this paper, we propose an AVSS framework that learns audio-visual joint feature representations from lip images, dense optical flow 
and audio stream with a novel cross-modal fusion. 
Extensive experiments show that compared to previous methods,
our cross-modal fusion based approach can exploit the semantic correlations between audio and visual stream, effectively, 
and we can also strengthen the robustness of visual representation through optical flow. 


\vspace{-10pt}
\section{Acknowledgment}

\vspace{-5pt}

This work was jointly supported by grants 61971352 \& 61862043 \& 61773397 \& 61801392 
approved by National Natural Science Foundation of China. The key grant 20204BC\\J22011 
approved by Natural Science Foundation of Jiangxi Province in China, the equipment pre 
research project 614000\\10115, as well as the National Science Foundation for Young 
Scientists of China 61801392.

\vfill\pagebreak


\bibliographystyle{IEEEbib}
\bibliography{refs}

\end{document}